\documentclass[a4paper,12pt,draft]{article}

\usepackage[a4paper]{geometry}
\usepackage[utf8]{inputenc}
\usepackage{hyperref}
\usepackage{float}
\usepackage{amssymb}
\usepackage{amsmath}
\usepackage{amsthm}
\usepackage{url}
\usepackage{amsfonts}
\usepackage{graphicx, float, tabularx}
\usepackage{color}
\usepackage{enumerate}
\usepackage[american]{babel}
\usepackage{stackrel}
\usepackage[numbers,sort,compress]{natbib}
\usepackage[normalem]{ulem}

\newcommand{\be}{\begin{equation}}
\newcommand{\ee}{\end{equation}}
\newcommand{\bea}{\begin{eqnarray}}
\newcommand{\eea}{\end{eqnarray}}

\renewcommand{\frac}[2]{{{\displaystyle #1}\over{\displaystyle #2}}}

\usepackage{amsmath}
\usepackage{latexsym}
\usepackage{color}

\include{amslatex}

\numberwithin{equation}{section}

\author{Ricardo Gallego Torrom\'e$^{a}$\thanks{E-mail: \texttt{rigato39@gmail.com}}, \and  Piero Nicolini$^{a,b}$\thanks{E-mail: \texttt{nicolini@fias.uni-frankfurt.de}}
\\[1ex]
\small $^a$ Frankfurt Institute for Advanced Studies (FIAS)\\[-0.5ex]
\small Ruth-Moufang-Str.~1, D-60438 Frankfurt am Main, Germany\\[1ex]
\small $^b$ Institut f\"{u}r Theoretische Physik, Goethe-Universit\"{a}t Frankfurt am Main\\[-0.5ex]
\small Max-von-Laue-Str.~1, D-60438 Frankfurt am Main, Germany\\[1ex]
}
\date{}
\title{Theories with maximal 
acceleration}

\begin{document}

\maketitle

\vspace{0.1cm}



\begin{abstract}
Maximal accelerations are related to the existence of a minimal
time for a given physical system. Such a minimal time can be either an
intrinsic time scale of the system or connected to a quantum gravity
induced ultraviolet cut off. In this paper we pedagogically introduce the
four major formulations for kinematics accounting for a maximal
acceleration. Some phenomenological repercussion are offered as hints for
future investigations.
\end{abstract}

\newpage

\tableofcontents{}

\section{Introduction}

The search of a consistent unified framework incorporating quantum theory and gravity has been a fundamental issue for the development of theoretical physics. After many discussions and research, it seems natural to think that such a theory should  come together with a modification of fundamental notions of physics. One of such notions could be the spacetime arena, the dynamical framework where physical description are setup.

In this context, the existence of a maximal acceleration has been demonstrated in various dynamical theories of quantum gravity, as in covariant loop quantum gravity \cite{RovelliVidotto} and in string theory \cite{ParentaniPotting}. Generally, the existence of a minimal scale in quantun gravitational models is related with the existence of an universal maximal acceleration
\cite{Kothawala, Bruneton}. Then one could wonder if these dynamical effects can be implemented in a modification of the spacetime structure itself, namely, in the form of a new  {\it kinematical geometry of maximal acceleration}.

  The notion of maximal acceleration appeared also in non-linear theories of classical electrodynamics \cite{Schuller}, in relation with the generalized uncertainty principle \cite{CapozzielloLambiaseScarpetta} and in theories addressing the problem of radiation-reaction \cite{Ricardo 2015,Ricardo 2017}, just to mention some examples. This suggests that there must exist a general kinematical formalism to accommodate a maximal acceleration in different theories.

Indeed, during the last decades there has been steady interest on the hypothesis of a maximal acceleration in Nature. By this we mean the general idea that the proper accelerations of test particles are bounded with respect to a given spacetime structure. The origin of this idea can be traced back to the foundational work of E. Caianiello \cite{Caianiello, Caianiello 1984} and H. E. Brandt \cite{Brandt1983}. In particular, Caianiello and his collaborators showed many interesting consequences of the existence of a maximal acceleration in several areas of theoretical physics. Among the suggested consequences, there is the modification of the  behaviour of singularities in cosmological models \cite{CaianielloGasperiniScarpetta}, in black hole solutions \cite{FeoliLambiasePapiniScarpetta}, showing the absence of absolute collapse (collapse to a point) of an extended gravitational body and the possibility of regularization of perturbative quantum field theory, specially for the effects on the structure of the propagators that maximal acceleration has \cite{NesterenkoFeoliLambiaseScarpetta}. These examples  illustrate the relevance of the kinematical approach to maximal acceleration and provide additional motivation for a thorough investigation of the idea.

\subsection{Arguments in favour of maximal proper acceleration}
There is a simple heuristic argument in favour of the existence of  maximal proper accelerations. Let us consider a physical system with a minimal time $\delta \tau$. The latter can be a characteristic time scale of the system or due to a natural ultraviolet cut off emerging from the background spacetime at Planckian energies \cite{Sprenger,Sabine}. If the system is relativistic, the lapse of time is associated to a proper time. Then in this case, the change in speed that the system can have is bounded by the speed of light divided by $\delta \tau$,
\begin{align}
|h(a,a)|\leq \frac{c^2}{(\delta \tau)^2},
\label{proper acceleration with tau}
\end{align}
where $h$ is a Lorentzian metric background with signature $(1,-1,-1,-1)$ and $a$ is the $4$-acceleration.
This argument is a direct generalization of Caldirola's argument \cite{Caldirola1981} from his theory of maximal acceleration for the extended electron \cite{Caldirola1956}.

Note that the above argument is substantiated on the existence of the fundamental time lapse $\delta{\tau}$ and such lapse could depend on the specific interaction producing the acceleration or the characteristic of the accelerated system. Paraphrasing A. Feoli \cite{Feoli}, there could be several  {\it maximal accelerations}. In the case that such maximal accelerations are associated not to the specific system, but to the interactions, we can speak of maximal proper accelerations in electrodynamics, in quantum gravity or for extended objects dynamics, to put some examples.
Indeed, the maximal acceleration could depend on the mass of the system \cite{Caianiello,Ricardo 2015,Schuller} or it could be universally defined \cite{Brandt1983, ParentaniPotting, RovelliVidotto}, for instance when associated with the Planck scale.

From the above discussion in turns out that the existence of an universal maximal acceleration is only consistent if there is an universal hierarchy in the accelerations and an adequate scaling of the maximal acceleration with the mass of the system.

Apart from this argument, the existence of an universal maximal acceleration has the following consequence. If the weak equivalence principle holds good, the existence of such scale implies the existence of a maximal gravitational field. Thus if the invariant expression of the gravitational field is the Ricci curvature, then the existence of the maximal acceleration implies bounded Ricci curvature.

\subsection{Examples of theories with maximal acceleration: a quick overview}
The first instance where the idea of proper maximal acceleration appeared was in the work of E. Caianiello \cite{Caianiello}. In his theory, each quantum particle of mass $m$ has heuristically associated a proper maximal acceleration whose modulus is given by
\begin{align}
a_{\mathrm{max}}=\,\frac{\mu \,c^2}{m\,\lambda},
\label{maximal acceleration for Caianiello 1}
\end{align}
where $\lambda$ is a quantity with dimension of length and $\mu$ a quantity with dimensions of mass, the interpretation of which is provided by Caianiello's theory \cite{Caianiello}. A similar expression is also derivable by direct application of Heisenberg's uncertainty principle \cite{Caianiello 1984}.

A different class of relations are obtained in models of quantum gravity. Thus for universal acceleration associated with the Planck scale, the expressions of the maximal acceleration\footnote{If the maximal acceleration in question depends on the characteristics of the systems, like mass or charge, it will be denoted by $a_{\mathrm{max}}$; if the maximal acceleration is an universal constant, independent of the system, it will be denoted by $A_{\mathrm{max}}$.} are of the form \cite{Brandt1983, ParentaniPotting, RovelliVidotto}
\begin{align}
A_{\mathrm{max}}=\,2\,\pi\,\alpha \,\left(\frac{c^7}{\hbar G}\right)^{1/2}
\label{maximal acceleration for gravity}
\end{align}
where $\alpha$ is a constant of order $1$. This maximal acceleration is of order $A_{\mathrm{max}}\sim 10^{52} m/s^2$.

In classical charged particle electrodynamics, some modifications of the Lorentz force equation imply that the maximal acceleration is bounded by an  expression of the form  \cite{Ricardo 2017}
\begin{align}
a_{\mathrm{max}}\sim m/q^2,
\label{Gallego Maximal acceleration}
\end{align}
where $m\neq 0$ is the mass of the particle and $q\neq 0$ its charge. In \cite{Ricardo 2017} this bound is obtained from the modified Lorentz-Dirac equation,
\begin{align}
m\,\ddot{x}=\,q \,F^\mu\,_\nu \dot{x}^\nu-\,\frac{2}{3}q^2\,h_{\rho\sigma}\,\ddot{x}^\rho\ddot{x}^\sigma\,\dot{x}^\mu
\label{modified Lorentz-Dirac equation}
\end{align}
in the regime where the modulus of the proper acceleration is much smaller than the maximal acceleration, where $h$ is a Lorentzian metric. The non-linear terms of the modification implies the condition
\begin{align*}
m^2\,|a|^2=\,|F_L|^2\left(1-\left(\frac{2}{3}\left(\frac{q}{m}\right)^2\right)^2\,|F_L|^2\right),
\end{align*}
where $|F_L|$ is the modulus of the external Lorentz force $F^\mu_L$. Then the assumption $F_L=\,ma+\,\textrm{higher order terms}$ and that $F^\mu_L$ must be spacelike, implies the condition \eqref{Gallego Maximal acceleration} in the regime where the acceleration is very small compared with $a_{\mathrm{max}}$. The bound \eqref{Gallego Maximal acceleration} coincides with the expression for the maximal acceleration found in Caldirola's theory of the electron \cite{Caldirola1981}.

Another example where maximal acceleration appears is in Born-Infeld non-linear theory of electrodynamics. In this case, the maximal acceleration is given by the expression \cite{Schuller}
\begin{align}
a_{\mathrm{max}}=\,\frac{q}{m} b^{-1},
\label{Born Infeld maximal acceleration}
\end{align}
where $b$ is the coupling of the Born-Infeld theory, a constant independent of the particle \cite{BornInfeld}.

Note the consistency of the limit $q\to 0$ in the expression \eqref{modified Lorentz-Dirac equation}: when the charge is neutral, there is no acceleration under an external electromagnetic field. Thus the maximal acceleration in such situations must be zero. Similarly, there is consistency with the limit $b\to 0$ in \eqref{Born Infeld maximal acceleration}.

We also observe that the different expressions above for the maximal accelerations are different and more importantly, have a different dependence on the mass-charge $(m,q)$ pair in different ways. This observation can be the basis for possible test of different models of point electrodynamics \cite{Ricardo 2017}.

\subsection{Aim and scope of the present work}
The present article has the purpose to give a critical and short overview of several relevant theories of maximal acceleration. It is well known that Lorentzian geometry and general relativity do not contain a maximal proper acceleration in their kinematical formalism. Thus if proper acceleration is bounded by a dynamical mechanism and such mechanism is of universal character, it could imply a modification of the kinematical theory itself \cite{Caianiello,  Brandt1983, Schuller, Ricardo 2015}. We will discuss these kinematical theories of maximal acceleration in this paper.

 It is not the aim of this work to offer an comprehensive overview of the investigations in the last decades concerning maximal acceleration. We focus the attention on  a particular argument line, namely, the attempts to find a complete and consistent kinematical theory with maximal proper acceleration. For other accounts  on maximal acceleration, the reader can have a look at \cite{Feoli,Toller2006}, for instance.   We will also not discuss the current status for the experimental search of maximal acceleration. Several proposals explore the transverse doppler effect and the corresponding bounds on the value of maximal acceleration obtained by application of Mössbauer spectroscopy. The interested reader could find further information and developments in\cite{Friedman, Friedman et al., Potzel}.
 Other proposal to test the existence of maximal proper acceleration(s) include deviations for the relativistic Thomas' precession law \cite{Schuller 2}.

\subsection{Notation}

In the next sections, we will adopt the following notation and symbols.
The four dimensional manifold will be indicated by $M$, while $M_4$ stands for a manifold diffeomorphic to $\mathbb{R}^4$. The symbol $TM$ indicates the tangent bundle of $M$ and $T^*M$ the co-tangent bundle. A generic Lorentzian structure signature $(1,-1,-1,-1)$ will be $(M, h)$, while the particular case of the Minkowski metric will be denoted by $\eta$. Greek indices run from $0$ to $3$. Equal up and down indices will be understood as contracted. If a local coordinate system for $M_4$ is given by the local coordinate functions $\{x^\mu,\,\mu=0,1,2,3\}$, then the associated local coordinates on $TM$ are given by the functions  metric in local coordinates on $TM$ given by $x^A\equiv (x^\mu, \frac{\hbar}{mc^2}\,\dot{x}^\mu)$. The acceleration with respect to $h$ will be denoted by $a^2$, namely, $a^2=\,h(D_{\dot{x}} \dot{x}, D_{\dot{x}}\dot{x})$, where $D$ is the covariant derivative of the Levi-Civita connection of $h$.

\section{Kinematic theories with maximal acceleration}

In this section we describe some of the kinematical theories of maximal acceleration, including some critical remarks on them.

\subsection{Caianiello's theory of maximal acceleration}

E. Caianiello introduced the idea of a maximal proper acceleration \cite{Caianiello} as a natural consequence of his geometric formulation of quantum mechanics. The geometrization of quantum mechanics proposed by Caianiello is based on a metric structure defined on the co-tangent space $T^*M_4$. If $\eta$ is the Minkowski metric on $M_4$, then there is a natural metric on $TM_4$,
\begin{align}
g_s=\,\eta\oplus_\alpha \eta^* .
\label{metric of maximal acceleration Caianiello 1}
\end{align}
Here $\alpha$ is a constant related with the value of the maximal acceleration, $\eta^*$ is the metric acting on the fiber space $\pi^{-1}(x)$ and $\oplus_\alpha$ means the {\it weighted direct sum operation}. In local coordinates, the metric from Caianiello on $T^*M_4$ i
s given by the line element \cite{Caianiello}
\begin{align}
c^2 \,ds^2=\,c^2\,dt^2- d\vec{x}^2\,+\frac{\hbar^2}{\mu^4 \,c^4}\left[\frac{1}{c^2} d E^2 - d\vec{p}^2\right],
\label{metric of maximal acceleration Caianiello 2}
\end{align}
where $\mu$ is a constant indicating a characteristic mass of the system.

The argument for the maximal acceleration from Caianiello follows from the criteria that for a massive particle the proper acceleration must be a spacelike four vector with respect to the metric \eqref{metric of maximal acceleration Caianiello 2},
\begin{align}
c^2-\,\vec{v}^2+\frac{\hbar^2}{\mu^4 \,c^4}\left[\frac{1}{c^2} \left(\frac{d E}{dt}\right)^2 - \left(\frac{d\vec{p}}{dt}\right)^2\right]\geq 0.
\end{align}
The evaluation of the left hand side using special relativity implies the bound
\begin{align*}
c^2-\,\vec{v}^2\left[1-\,\frac{\hbar^2}{\mu^4 \,c^4}\, \frac{m^2\vec{a}^2\,c^4}{(c^2-\,\vec{v}^2)^{3}}\right]\geq 0.
\end{align*}
Therefore, the proper acceleration $\vec{a}^2$ is bounded by the expression \eqref{maximal acceleration for Caianiello 1}.

Several direct implications of the theory  were discussed by Caianiello. Perhaps, the most relevant consequence is the fact that the existence of a maximal acceleration avoids the total collapse of a black hole \cite{Caianiello}.

There is another argument, also proposed by Caianiello (see also \cite{Brandt1983}) that shows the need of a maximal proper acceleration. This second  argument is based on Heisenberg uncertainty principle \cite{Caianiello 1984}. The starting point is the relation
\begin{align*}
\Delta E \Delta f(t)\geq \,\frac{\hbar}{2}\, \left| \frac{df}{dt}\right|,
\end{align*}
where here $t$ is an arbitrary time parameter.
If $\Delta E\leq \, E$, $f=v$ and the relativistic constraint $\Delta v\leq c$ holds, then from the relativistic relation $E=mc^2$ and applied to the coordinate system where the particle is at rest, we have that the proper acceleration $a$ must be bounded by a maximal value given by
\begin{align}
a_{\mathrm{max}}=\,2\,\frac{mc^3}{\hbar}.
\label{Maximal acceleration of Caianiello 2}
\end{align}
Compared with the expression \eqref{maximal acceleration for Caianiello 1}, this value of the maximal acceleration depends on the mass $m$ of the particle, instead of the characteristic mass $\mu$. This derivation of the value of maximal proper acceleration do not depend upon the existence of a quantum of time $\delta \tau$ or characteristic length $\lambda$.

A different formulation of the theory of maximal acceleration of Caianiello is found in \cite{CaianielloFeoliGasperiniScarpetta}. There, the flat spacetime manifold $M_4$ is embedded in the tangent space $TM_4$. The tangent space has a natural Sasaki-type metric of the form
\begin{align}
g_S=\,\eta\oplus_\alpha \eta .
\label{metric of maximal acceleration Caianiello 3}
\end{align}
In this framework, it is postulated the proper time measure by physical clocks is given by the line element
\begin{align}
d\tilde{s}^2=\,g_{AB}\,dx^A\,dx^B=\,\eta_{\mu\nu}\,dx^\mu \,dx^\nu +\,\frac{\hbar^2}{4\,m^2c^6}\,\eta_{\mu\nu}\, d\dot{x}^\mu\,d\dot{x}^\nu.
\label{metric of maximal acceleration Caianiello 4}
\end{align}
Physical particles have associated a physical vector velocity field. Then for a massive particle, the induced proper time element in $M_4$ is given by \cite{CaianielloFeoliGasperiniScarpetta}
\begin{align}
d\tau^2 =\,ds^2\left(1-\frac{\hbar^2}{4\,m^2c^6}\,|\eta_{\mu\nu}\,\ddot{x}^\mu \,\ddot{x}^\nu|\right).
\label{metric of maximal acceleration Caianiello 5}
\end{align}
In this expression, derivatives are taken with respect to $ds$, the proper time of the Minkowski metric $\eta$.
 It follows that the value for the maximal acceleration given by Caianiello's formula \eqref{Maximal acceleration of Caianiello 2}. The requirement that the proper time of a massive particle is positive or zero is translated now to the condition
\begin{align*}
1-\frac{|\eta_{\mu\nu}\,\ddot{x}^\mu \,\ddot{x}^\nu|}{m^2}\geq 0,
\end{align*}
which implies the existence of a bound for the proper acceleration $a^2=\,\eta_{\mu\nu}\,\ddot{x}^\mu \,\ddot{x}^\nu$ given by the maximal acceleration \eqref{maximal acceleration for Caianiello 1}.


Let us remark that the maximal accelerations \eqref{metric of maximal acceleration Caianiello 1} and \eqref{metric of maximal acceleration Caianiello 3} are different
and henceforth, provide different theories of maximal acceleration.

An interesting example discussed in \cite{CaianielloFeoliGasperiniScarpetta} is the case of the modification of the Rindler metric. In standard coordinates, the Rindler $1+1$ spacetime metric is
\begin{align*}
ds^2=\,\chi^2 \,d\kappa^2-\,d\chi^2,
\end{align*}
with $-\infty \, <\kappa <\, +\infty$, $0<\,\chi < \, +\infty$. This metric is modified in Caianiello's theory to the form
\begin{align}
ds^2=\,(\chi^2-m^{-2}) \,d\kappa^2-\,d\chi^2.
\end{align}
This is not a flat metric; the scalar curvature is
\begin{align}
R=\,-\frac{2}{m^2}(\chi^2-m^{-2})^{-2},
\end{align}
which has a singularity at the modified Rindler horizon $\chi=m^{-1}$. This is an example of how the existence of a maximal acceleration modifies the causal structure of the spacetime and also prevents the system from collapsing to a single point. It was shown that the existence of the singularity at the horizon implies the emergence of an effective repulsive force, that avoids any test particle to penetrate the interior and the system to collapse \cite{CaianielloFeoliGasperiniScarpetta}.

Caianiello's theory has many other significant consequences. Among them there is the appearance of a deflection mechanism in extended object cosmology, that prevents the formation of singularities at the cosmological level \cite{CaianielloGasperiniScarpetta}, the formation of a shell out of the horizon in the quantum geometry modifications  Schwarzschild black body solution \cite{FeoliLambiasePapiniScarpetta} and the Kerr spacetime type solution \cite{BozzaFeoliLambiasePapiniScarpetta2}, violations of the weak equivalence principle \cite{BozzaLambiasePapiniScarpetta3}. All these modifications vanish in the limit $\hbar \to 0$. In this sense, these corrections  have a {\it quantum origin}.

Another type of consequence comes from the application to the regularization of quantum field theories \cite{NesterenkoFeoliLambiaseScarpetta}. This is because the Lagrangian of particle moving with maximal proper acceleration has associated a Green function with a higher order momentum in the denominator. It is plausible that such quantum field theories could be ultra-violet finite and at the same time, being compatible with special relativity.

\subsection{A critical view on Caianiello's theory}

Despite its far reaching consequences, the formulation of Caianiello's theory has several dramatic limitations. Let us start by considering the theory of maximal proper acceleration geometry developed in \cite{CaianielloFeoliGasperiniScarpetta}, in particular the expression in local coordinates for the metric of maximal acceleration \eqref{metric of maximal acceleration Caianiello 5}. This expression is not general covariant, which could be problematic for any theory aimed to embrace also gravity. The root of the problem is found in the definition of the Sasaki type metric \eqref{metric of maximal acceleration Caianiello 4}.  
 
In order to have a covariant Sasaki metric, it is necessary to introduce a non-linear connection. Let us consider the element of the form \cite{Brandt1989, GallegoTorrome}  for a generic Lorentzian metric $h$ on $M$, generalizing the construction starting from the Minkowski metric $\eta$,
\begin{align}
g_s=\,g_{AB}\,dx^A\otimes dx^B = \,h_{\mu \nu} dx^\mu \otimes dx^\nu + h_{\mu \nu}\Big( {\delta \dot{x}^{\mu}}\otimes {\delta \dot{x}^{\nu}} \Big).
\label{Sasaki type metric}
\end{align}
In this expression the $1$-forms $\delta{\dot{x}}^\mu$ contain corrections due to the non-linear connection that makes the expression properly covariant under local coordinate changes.
An introduction to these geometric notions can be found in \cite{GallegoTorrome} and a comprehensible treatments in \cite{BCS, MironHrimiucShimadaSabau:2002}. The notion of non-linear connection is similar to the usual notion of affine connection, that allows for covariant differential of {\it sections} of vector bundles\footnote{Given a vector bundle $\mathcal{E}$ with canonical projection $\pi:\mathcal{E}\to M$ on the base manifold $M$, a {\it section} $S$ is a map $S:M\to \mathcal{E}$ such that $\pi\circ S$ is the identity map $Id:M\to M$.}.

 The induced proper time element along congruences of world lines on $M$ of the metric \eqref{Sasaki type metric} is of the form
\begin{align}
d\tau^2 =\,\left(1-\frac{\left|(D_{\dot{x}}\,\dot{x})^{\sigma}(D_{\dot{x}}\,\dot{x})_{\sigma}\right|}{a^2_{max}}\right)\,ds^2 .
\label{covariant formulation of the proper time metric of maximal acceleration}
\end{align} Note that for $\left|(D_{\dot{x}}\,\dot{x})^{\sigma}(D_{\dot{x}}\,\dot{x})_{\sigma}\right|<a^2_{\mathrm{max}}$, this expression is well defined.
The covariant derivative here is the one associated with the Levi-Civita connection of the metric $h$, although it could be in principle associated to any affine connection on $M$. This construction is valid for an arbitrary Lorentzian manifold $(M,h)$, not only for Minkowski spacetime. Also, we did not specify the value of the maximal acceleration parameter $a_{\mathrm{max}}$.

 The second main problem that we find with the above construction, perhaps deeper than the previous one, is that Caianiello's maximal acceleration geometry \eqref{metric of maximal acceleration Caianiello 4} and the Sasaki type metric \eqref{Sasaki type metric} need of an underlying Lorentzian structure $(M,h)$. The  proper time can be either constructed from the underlying Lorenzian spacetime $(M,h)$ or from the metric of maximal acceleration \eqref{metric of maximal acceleration Caianiello 4}. Therefore, the existence of more than one  metric structure implies a dichotomy. If $(M,g)$ is the physical structure (by assumption, it determines the physical proper time), then it is unclear how one can obtain the Lorentzian structure $(M,h)$ by an operational method.

\subsection{Brandt's differential geometric approach to maximal acceleration}

In 1983, H. E. Brandt provided several heuristic arguments for the existence of a maximal universal acceleration \cite{Brandt1983}. One of these arguments started considering Sakharov maximal temperature \cite{Sakharov 1966}. The maximal temperature that a system can have in equilibrium with black body radiation turns to be given by
\begin{align}
T_{max}\equiv\,\frac{\alpha}{k}\sqrt{\frac{c^5 \,\hbar}{G}},
\label{maximal temperature}
\end{align}
where $k$ is the Boltzmann's constant and $\alpha$ a dimensional number of order unity. On the other hand, for an accelerated frame with proper acceleration $a$, the co-moving observers experiment a thermal bath in vacuum with temperature \cite{Davis 1975, Unruh 1976}
\begin{align}
T=\,\frac{\hbar\,a}{2\,\pi\,k\,c}.
\label{temperature of the thermal bath}
\end{align}
Therefore, the maximal acceleration that a system can have is given by the expression \eqref{maximal acceleration for gravity}. Remarkably, Caianello and Landi argued how from the expression of maximal acceleration in Caianiello's theory one can re-derive Sakharov's maximal temperature \cite{CaianielloLandi}.

 Motivated by the above argument, Brandt developed a geometric approach to systems where the proper acceleration is bounded by a maximal acceleration  $A_{\mathrm{max}}$. If for a world line of a physical particle  $a\leq A_{\mathrm{max}}$, then
\begin{align*}
\left|(D_{\dot{x}}\,\dot{x})^{\sigma}(D_{\dot{x}}\,\dot{x})_{\sigma}\right|\leq A^2_{\mathrm{max}}.
\end{align*}
This relation was interpreted in \cite{Brandt1989} in terms of the positiveness of the bilinear form
\begin{align}
h_{\mu\nu}\,dx^\mu\,dx^\nu+\,\frac{1}{A^2_{max}}\,h_{\mu\nu}\,\left(d\dot{x}^\mu+\,\Gamma^\mu_{\alpha\beta}\,\dot{x}^\alpha\,dx^\beta\right)\,
\left(d\dot{x}^\nu+\,\Gamma^{\nu}_{\delta\gamma}\,\dot{x}^\delta\,dx^\gamma\right),
\label{Brandt form}
\end{align}
where here $\Gamma^\mu_{\alpha\delta}$ are the connection coefficients of the affine connection $D$, usually taken the Levi-Civita connection of $h$ and it is assumed that $A_{\mathrm{max}}>0$, that is, $|a|<A_{\mathrm{max}}$.
The expression \eqref{Brandt form} is the bilinear, symmetric form defined on $TM$. Thus if one also assumes that this form is non-degenerate, it defines a metric on $TM$,
\begin{align}
G=\,G_{AB}\,du^A\,du^B,
\label{Brandt metric}
\end{align}
where $u^A=\,(x^\mu,\dot{x}^\mu)$ are natural coordinates in $TM$.
In natural coordinates, the metric $G$ has the following matrix components,
\begin{align}
G_{AB}= \,
\left( \begin{array}{cc}
h_{\mu\nu}+\,\frac{1}{A^2_{max}}\,\Gamma^\alpha_{\lambda\mu}\,h_{\alpha\beta}\,\Gamma^{\beta}_{\delta\nu}\,\dot{x}^\lambda \dot{x}^\delta & \frac{1}{A_{\mathrm{max}}}\,h_{\alpha \nu}\,\Gamma^\alpha_{\delta \mu}\,\dot{x}^\delta\\
 & \\
\frac{1}{A_{\mathrm{max}}}\,h_{\alpha \mu}\,\Gamma^\alpha_{\delta \nu}\,\dot{x}^\delta & h_{\mu\nu}
\end{array}\right) .
\label{Brandt metric components}
\end{align}
Except for the value of the proposed maximal acceleration, the metric \eqref{Brandt metric} coincides with the so-called Sasaki type metric discussed in reference \cite{GallegoTorrome}. Also note that in the case  the metric $h$ is the Minkowski metric $\eta$, then \eqref{Brandt metric}  metric coincides with Caianiello's metric \eqref{metric of maximal acceleration Caianiello 3}. However, Brandt's theory is a manifestly general covariant theory.

Brandt's theory is based upon the assumption that the physical metric, the one that is testable  by experiments performed by macroscopic observers, is given by \eqref{Brandt metric}. Brandt also considered the geometric theory for the metric \eqref{Brandt metric} and formulated the corresponding field equations, generalizations from Einstein equations of general relativity. Also, an interpretation of the metric \eqref{Brandt metric} as a Kaluza-Klein type metric was explored \cite{Brandt1989}. Brandt formulated field theories in this framework during the 90', showing that many of the fundamental concepts of modern field theory could be formulated in his theory (see for instance \cite{Brandt1998} and references there).

However, the starting point in the construction of the metric \eqref{Brandt metric} is the spacetime metric $h$ defined on $M$. By the assumptions of the theory, $h$ is not the metric that should be reconstructed by operational measurements. Indeed, one needs this metric $h$ as a back-ground structure defined on $M$, prior to the construction of $G_{AB}$. Thus, analogously
 as in Caianiello's theory, one has in Brandt's theory a dichotomy between $(M, h)$ and $(TM, G)$, since both structures can be used to define observables, for instance, proper time for a generic world line $x:I\to M$. It is unclear what is the physical meaning of $h$ in Brandt's theory.

\subsection{Schuller's algebraic formulation of Born-Infeld theory}
 F. P. Schuller formulated a kinematical theory of maximal acceleration in \cite{Schuller},  motivated by Born-Infeld theory \cite{BornInfeld}. It is based on a pseudo-complex extension of Lorentzian geometry as follows. The commutative ring of pseudo-complex numbers is defined by the set
 \begin{align*}
 \mathbb{P}=\,\{a\,+I\,b\,|\,a,b\in\,\mathbb{R}\}
 \end{align*}
equipped with addition and multiplication laws induced by those on $\mathbb{R}$ and such that $I$ is a pseudo-complex structure, namely, the relation $I^2=1$ holds. There is a matrix representation of $\mathbb{P}$. Thus if $u=a\,1+I \,b\in \mathbb{P}$, then the matrix representation is such that
\begin{align*}
1\equiv \,
\left( \begin{array}{cc}
1 & 0\\
0 & 1
\end{array}\right),\quad \quad I \equiv\, \left( \begin{array}{cc}
0 & 1\\
1 & 0
\end{array}\right),
\end{align*}
where the operations on $\mathbb{P}$ correspond to the matrix operations.

 A $\mathbb{P}$-module is an algebraic structure analogous to a vector space, but where the coefficients in the linear combinations are taken with respect to a ring $\mathbb{P}$, instead than to a numerical field, for instance the real numbers $\mathbb{R}$.
It can be shown that fundamental constructions, like $\mathbb{P}$-extension of a Lie algebra\footnote{If $L$ is a real Lie algebra, a $\mathbb{P}$-extension is an algebra of the form $\alpha_1 + \alpha_2 I$, where $\alpha_1,\,\alpha_2\in \,L$ and the bracket operation is extended by linearity.}, exponential map, etc... carries over the $\mathbb{P}$-module representations in a closer form to the real Lie algebra theory over real vector spaces \cite{Schuller}.
In particular, the pseudo-complex Lorentz group
\begin{align}
O_{\mathbb{P}}(1,3)\equiv \{\Lambda \in \,\mathrm{Mat}(n,\mathbb{P}) \,s.t.\,\Lambda^\top\,\eta \Lambda=\,\eta,\,\det \Lambda=1\}
\end{align}
will play a relevant role in the generalized kinematics.
At the level of the Lie algebra, one has the identity
\begin{align}
so_{\mathbb{P}}\cong\, so_{\mathbb{R}}(1,3)\oplus \,so_{\mathbb{R}}(1,3).
\end{align}
The Lie group $O_{\mathbb{P}}(1,3)$ leaves invariant the bilinear form
\begin{align}
\eta_{\mathbb{P}}:TV_{\mathbb{P}} \otimes TV_{\mathbb{P}} \to \,\mathbb{P},
\label{pseudoconvex lift}
\end{align}
induced from the Minkowski metric defined on $M_4$, where here $V_{\mathbb{P}}$ is the pseudo-complexification of the vector space $M_4\cong \mathbb{R}^4$ and $TV_{\mathbb{P}}$ is its tangent space. Thus one can see that $V_{\mathbb{P}}\cong T M_4$ by comparing the corresponding real dimensions. In terms of the representation of $V_{\mathbb{P}}\cong T M_4$, a point $v \in V_{\mathbb{P}}$ is identified with $(x, \dot{x})$. Therefore, a generic point of $TV_{\mathbb{P}}$ is identified with $(\dot{x},\ddot{x})\in\, TT_{(x,\dot{x})}M_4.$

The bilinear form $\eta_{\mathbb{P}}$ defines the metric of maximal acceleration through two additional postulates:
\begin{itemize}

\item For physical orbits on $TTM_4$, the metric $\eta_{\mathbb{P}}(\dot{X},\dot{X})\geq 0$,

\item The proper time of a physical orbit $X:[a,b]\to TM_4$ is given by the proper time of $\eta_{\mathbb{P}}$.

\end{itemize}
It can be shown that this theory implies a consistent bound of the proper maximal acceleration with respect to $\eta$ \cite{Schuller}. Indeed, the line element associated to this "metric of maximal acceleration is of the form
\begin{align}
dw^2=\,\left(1-\frac{a^2}{a^2_{max}}\right)\,ds^2.
\label{Schuller maximal acceleration element}
\end{align}
We observe that the structures \eqref{pseudoconvex lift} and \eqref{Schuller maximal acceleration element} in Schuller's theory are equivalent to the corresponding structures \eqref{metric of maximal acceleration Caianiello 3} and \eqref{metric of maximal acceleration Caianiello 5} in Caianiello's theory. However, note that in Schuller's theory $a_{\mathrm{max}}$ is not fixed by Caianiello's maximal acceleration \eqref{Maximal acceleration of Caianiello 2}. Instead $a_{\mathrm{max}}$ is a free parameter with dimension of proper acceleration. In particular and since it was motivated by Born-Infeld theory, $a_{\mathrm{max}}$ could be given by the expression \eqref{Born Infeld maximal acceleration}.

Schuller's theory  can be implemented in curved spacetimes $(M,h)$, by a point-wise generalization on $M$ of the above algebraic construction. It is also consistent with a canonical quantization procedure. This contrasts with previous formulations of Caianiello's theory based on complex structures compatible with the connection associated with the metric \eqref{metric of maximal acceleration Caianiello 3}. Indeed, it was proved that under reasonable assumptions, Caianiello's metric \eqref{metric of maximal acceleration Caianiello 3} is not consistent with quantization, except if the metric $h$ is flat \cite{Schuller}. Related with this, it could be of relevance to investigate whether the pseudo complex General Relativity of Hess and Greiner \cite{Hess and Greiner 2009} provides a framework for curve spacetimes consistent with maximal acceleration.

 Schuller's theory is general covariant, independent of the local coordinate used to be formulated. However, the second problem discussed for Caianiello's theory is also present in Schuller's construction, since it depends upon an underlying Lorentzian structure $(M,h)$, an apparently ad hoc structure in the framework of metrics with maximal acceleration.

\subsection{An effective theory of metric geometries with maximal acceleration and jet geometry}

The notion of proper maximal acceleration is not necessarily linked with the quantum description of physical systems. Indeed, since the proper acceleration is defined as the derivative of the four-velocity with respect to the proper time along a given world line, it seems more appropriate to think the concept of maximal proper acceleration from a classical point of view.

It also seems clear that there are different notions of maximal proper acceleration and that such notions depend upon the system and the dynamics involved. Therefore, it is natural to leave un-specified the value of the maximal acceleration $a_{\mathrm{max}}$ in the search of classical geometric frameworks for maximal proper acceleration.

One of the fundamental features of Caianiello's, Brandt's and Schuller's theories is that the proper time depends not only on the instantaneous value of the speed with respect an inertial or free falling coordinate system, but also on the instantaneous four-acceleration. This is a violation of the so-called clock hypothesis in the theory of relativity \cite{Rindler}, when the problem of the description physical phenomena in an accelerated coordinate system is considered.

On the other hand, violations of the clock hypothesis are expected to happen in physical situations where radiation reaction effects are of relevance \cite{Mashhoon1990,Mashhoon2}. These situations are of particular relevance in electrodynamics, where the standard equation of motion of a point charged particle \cite{Dirac1938} present serious theoretical difficulties.

It is in the above context that the theory developed in \cite{Ricardo 2015} must be interpreted. The fundamental idea is that in the treatment of certain physical processes, the clock hypothesis could be violated, according to B. Mashhoon \cite{Mashhoon1990,Mashhoon2}. Therefore, the natural generalization of the proper time must depend upon the acceleration. The  metric of maximal acceleration must be  a fundamental physical element. The fundamental idea of the theory is that the mathematical description of a metric of maximal acceleration of the type \eqref{metric of maximal acceleration Caianiello 5} is a geometric structure whose components live on the second jet bundle over $M$. This is the bundle determined by the set
\begin{align*}
J^2_0(M_4)\equiv \{(x,\dot{x},\ddot{x}),\,x:I\to M_4 \,\textrm{smooth},\, 0\in I\subset \mathbb{R}\},
 \end{align*}
 where the  coordinates of a given point $u\in\,J^2_0(M_4)$ are of the general form
 \begin{align*}
 (x,\dot{x},\ddot{x})=\,\left( x^\mu(s),\frac{dx^\mu(s)}{ds},\frac{d^2 x^\mu(s)}{ds^2}\right)
  \end{align*}
  and by the canonical projection $\pi_2:J^2_0\to M_4,\,(x,\dot{x},\ddot{x})\mapsto x$.
  
We recall here that jet theory is a framework to systematically deal  with Taylor expansions of
functions on manifolds and sections. Roughly speaking, the $k$-jet of a
function or section at a given point corresponds to a Taylor expansion up
to order $k$, without considering the remaining term of order $k$+1. For
instance, the $k$-jet of a curve $$\gamma:(-a,a)\to M$$ at the point
$\gamma(0)\in\, M$ is 
$$(\gamma^\mu (0), \dot{\gamma}^{\mu}
(0),\ddot{\gamma}^{\mu}(0)..,\gamma^{\mu(k)}(0)),$$ where $$\dot{\gamma}^\mu =
\,\frac{d\gamma^\mu}{ds},\quad \gamma^{\mu(k)}=\,\frac{d^k\gamma^\mu}{ds^k} $$ evaluated at the point $\gamma(0)$ and $s$
is the parameter of the curve. Different parameters determine different
jets, which are, nevertheless related by strict transformation rules. The
collection of possible jets, for instance, the collection of $k$-Taylor
expansions of curves on a manifold, can be fulled with differentiable
structure, which makes them manifolds. Furthermore, there are natural
projections that provide further structure to such manifolds, such as fiber bundles. 
Therefore, one can speak of jet bundles \cite{KolarMichorSlovak}.
  
 Let us consider the general case where the spacetime manifold is $M_4$. Then the {\it maximal acceleration metric} it is postulated to be
  determined by a map that associates to each {\it physical world line} $x:I\to M_4$ a smooth family of scalar products, one at each point of the curve,
\begin{align*}
\{g(\,^2x(t)):T_{x(t)}M\times T_{x(t)}M\to \mathbb{R},\,t\in I\}
\end{align*}
along the world line $x:I\to M$ whose components live on the second jet lift $^2x:I \to J^2_0(M)$. This family of scalar products can formally be expressed in a general way as
 \begin{align}
 g(\,^2x)=\,g^0(\,x,\dot{x}, \ddot{x})+\,g^1(x,\dot{x}, \ddot{x})\xi(x,\dot{x},\ddot{x},a^2_{max}),
  \label{perturabativeexpansion}
 \end{align}
where dot derivatives $\dot{x}$ are meant with respect to the proper parameter of the metric $g_0$.

 We can consider limits when $a_{\mathrm{max}}\to +\infty$ in the family of metrics $g(a_{\mathrm{max}})$. Then we require that metric obtained by the limit
\begin{align*}
\lim_{a^2_{max}\to +\infty}\,g(\,^2x)
\end{align*}
 is compatible with the clock hypothesis. We also assume that $\xi(x,\dot{x},\ddot{x},a^2_{max})$ is analytical in $1/a^2_{max}$ and has the form
\begin{align}
\xi(x,\dot{x},\ddot{x},a^2_{max})=\,\sum^{+\infty}_{n=1}\,\xi_n(x,\dot{x},\ddot{x})\,\left( \frac{1}{a^2_{max}}\right)^n.
\label{generaltypeofcorrections}
\end{align}
Then one can argue that
\begin{align*}
\lim_{a^2_{max}\to +\infty}\,g(\,^2x)=\,g^0(x,\dot{x},\ddot{x})
\end{align*}
and by compatibility with the clock hypothesis,
\begin{align*}
g^0(x,\dot{x},\ddot{x})\equiv\,g^0(x,\dot{x}).
\end{align*}
The form $g^0(x,\dot{x})$ is non-degenerate, since $g$ is non-degenerate. $g^0$ is also symmetric and bilinear. Therefore, $g^0$ is indeed a generalized Finsler metric \cite{GenFinRGT}. If we make the further assumption that $g^0=\,h$ is Lorentzian and we assume the identifications
\begin{align*}
g^0 =\,h,\quad  g^1 =\,h,\quad \xi_1=\,h(\ddot{x},\ddot{x})
\end{align*}
then we have a generalization of the line element \eqref{metric of maximal acceleration Caianiello 5}.
In this case, the metric of maximal acceleration \eqref{perturabativeexpansion} can be expressed as
\begin{align}
g_{\mu\nu}(\,^2x):=\,\left[1- \frac{\left| h(D_{\dot{x}}\dot{x}(t),
D_{\dot{x}}\dot{x}(t))\right|}{g^0(\dot{x},\dot{x})\,a^2 _{max}}\right]h_{\mu\nu}
\label{maximalaccelerationmetric2}
\end{align}
where the curve $x:I\to M$ is parameterized by the proper time parameter of $g$ in the limit when $a_{\mathrm{max}}$ tends to zero. In practical examples, this limit metric coincides with $h$.  One example where this procedure is consistent is the model for point charged particles discussed in \cite{Ricardo 2017}.

The main advantage of this formalism with respect to others kinematical theories of maximal acceleration rests on the fact that from the beginning, the theory is formulated in terms of the metric of maximal acceleration; the theory just discussed is based on the notion of maximal acceleration given by a family of metrics defied on the second jet. Thus, apparently, the above mentioned dichotomy in Caianiello, Brandt and Schuller's theories is resolved. The metric $h$ has only a formal definition and it is attached to $g$ and not the other way around.

 However, the theory discussed above is in principle an effective theory, since in principle the theory as it is formulated is limited to a range of validity when $a<<a_{\mathrm{max}}$. A deeper treatment is missing.

\section{Conclusions}
The idea of maximal acceleration and the associated modification of the spacetime geometry has been around for long time. Being a modification of the fundamental ideas of the theory of relativity, the theoretical consequences of maximal acceleration are sound. However, the current state of the art shows important deficiencies, probably the reason of why maximal acceleration has been not yet attracted enough attention. The theories of maximal acceleration are such that either they lack of a general covariant formulation or if such formulation exits \cite{Ricardo 2015}, then it has the limitation to be perturbative and with limited domain of validity, far from the region of maximal acceleration. Also, other formulations \cite{Caianiello, Brandt1989, Schuller, GallegoTorrome} are rooted on a pre-existent Lorentzian metric back-ground, with the conceptual consequences that this entails.
These problems, however, are likely to be technical problems.

There is also some relation between the idea of maximal acceleration in
physics and the conjecture of {\it maximal tension or force} \cite{Gibbons2002}. This conjecture asserts that in general relativity, there is a limit
to the tension or force that a physical system can feel, a limit given by
the expression
\begin{align*}
F\leq F_{\mathrm{max}} = \,\frac{c^4}{4 G},
\end{align*}
where $G$ is Newton's gravitational constant.
The factor 1/4 in this expression can vary, in the sense that it could
depend on the details of the dynamical system. For instance, in presence of
a cosmological term in some models of black-holes, this factor is instead
given by 1/9 \cite{BarrowGibbons2014}.

A general argument supporting this conjecture is based upon considerations
on black hole merging in general relativity and  a form of {\it cosmic censorship} \cite{Gibbons2002,
BarrowGibbons2014}. It is in any case, a classical argument.

The reader should note that the notions of maximal universal force and of
maximal acceleration are not equivalent in general relativity. In fact, the
mathematical structure of general relativity does not contain a notion of
universal maximal acceleration, since in such a framework, there is no a
minimal universal length scale. Thus if the mathematical framework is kept 
Lorentzian or pseudo-Riemannian,  the interplay between maximal acceleration and maximal forcethese implies the
introduction of new principles, like the aforementioned existence of a minimal length in
quantum gravity \cite{Gibbons2002}. Maximal acceleration is, however,
compatible also with classical models in the case of generalized geometric
frameworks, like maximal acceleration geometry \cite{Ricardo 2015}.

In a theory where the maximal acceleration could
depend upon the system itself, like in Born-Infeld dynamics, it is not necessary
to introduce quantum mechanics to make compatible maximal acceleration
with maximal universal force.
To the maximal force $c^4/4G$ corresponds the maximal acceleration
$c^4/4 G m$.

Similar related conjectures, like maximal power and maximal angular
momentum \cite{BarrowGibbons2014,BarrowGibbons2017} could be related
with maximal acceleration in analogous ways. All these conjectures are
likely to be seen also in a kinematical version, where maximal acceleration
is involved.

From the phenomenological point of view, it has been investigated possible phenomenological signatures for maximal acceleration \cite{Friedman et al., Potzel, Schuller 2}. However, these different results only provide lower bounds for the maximal acceleration. Another possibility has been discussed in \cite{Ricardo 2017}, linked with a new theory of classical electrodynamics. It was predicted the strict decrease of the maximal acceleration with the size (charge and mass) of the charge particle.

Interestingly the existence of a maximal acceleration could have repercussions also on the physics of evaporating black holes \cite{NiW11} and the related Unruh effect that has been at the center of a recent debate in the literature \cite{NiR11}.

  We have discussed only some aspects of the main idea of having a kinematical theory with maximal acceleration. It is to be expected that such a profound and simple idea of maximal acceleration could have further interesting consequences, both at the phenomenological and theoretical levels of physical description.

\subsection*{Acknowledgements}

The authors are grateful to John D. Barrow and Dawood Kothawala for correspondence and references.


\end{document}